# Structure and Anisotropic Properties of $BaFe_{2-x}Ni_xAs_2$ (x = 0, 1, 2) Single Crystals

A. S. Sefat, R. Jin, M. A. McGuire, B. C. Sales, D. Mandrus
*Oak Ridge National Laboratory, Oak Ridge, TN 37831, USA*

F. Ronning, E. D. Bauer
*Los Alamos National Laboratory, Los Alamos, NM 87545, USA*

Y. Mozharivskyj
*Chemistry Department, McMaster University, Hamilton, ON L8S 4M1, Canada*

## Abstract

The crystal structure, anisotropic electrical resistivity and magnetic susceptibility, as well as specific heat results from single crystals of $BaFe_2As_2$, $BaNi_2As_2$, and $BaFeNiAs_2$ are surveyed. $BaFe_2As_2$ properties demonstrate the equivalence of C(T), Fisher's $d(\chi T)/dT$, and $d\rho/dT$ results in determining the antiferromagnetic transition at $T_N$ = 132(1) K. $BaNi_2As_2$ shows a structural phase transition from a high-temperature tetragonal phase to a low-temperature triclinic ($P\bar{1}$) phase at $T_0$ = 131 K. The superconducting critical temperature for $BaNi_2As_2$ is well below $T_0$ and at $T_c$ = 0.69 K. $BaFeNiAs_2$ does not show any sign of superconductivity to 0.4 K and exhibits properties similar to $BaCo_2As_2$, a renormalized paramagnetic metal.

## 1. Introduction

The discovery of superconductivity in a variety of compounds with square lattice sheets of $Fe^{2+}$ or $Ni^{2+}$ in the structure has created great attention primarily due to their non-copper based origin. The structural parents of iron-based superconductors include those of ZrCuSiAs-type *R*FeAsO [1], $ThCr_2Si_2$-type $BaFe_2As_2$ [2], PbFCl-type LiFeAs [3], and PbO-type FeSe [4]. Like in the cuprates, the superconductivity in the Fe-based compounds is suggested to be unconventional as the electron-phonon coupling is thought to be too weak to account for the observed high critical temperatures [5-7]. Spin fluctuations derived from multiple *d* orbitals and itinerant electrons may mediate the pairing interaction in Fe-based superconductors, [8-12]. The Fe-based superconductors are low carrier density metals with separated Fermi surfaces and proximity to itinerant magnetism, including an itinerant spin-density-wave (SDW). The suppression of magnetism in the parent compounds is found to give way to superconductivity, although their possible coexistence has been suggested [9, 13-15]. The Ni-based superconductors have much lower critical temperatures and include those of LaNiAsO [16, 17], $SrNi_2As_2$ [18] and $BaNi_2As_2$ [19, 20]. LaNiAsO exhibits bulk superconductivity with $T_c \sim 2.75$ K and is likely a conventional Bardeen-Cooper-Schrieffer (BCS) superconductor [17, 21]. For Pb-flux grown $BaNi_2As_2$ crystals, a superconducting critical temperature at $T_c = 0.7$ K was found. $BaNi_2As_2$ gives a first order phase transition at $T_0 = 130$ K upon cooling, and at 137 K upon warming [19]. By analogy with $AFe_2As_2$ ($A$ = Ba, Sr, Ca) [22-25], this transition was tentatively identified in Ref. [19] as a structural transition from a high-temperature tetragonal to low-temperature orthorhombic phase, possibly related to a SDW magnetic transition. However, here we find much lower symmetry for the low temperature structure of $BaNi_2As_2$.

Due to the existing interest in the properties of $ThCr_2Si_2$-type $BaT_2As_2$ (T = transition metal) materials, pure, high-quality single crystalline samples of $BaFe_{2-x}Ni_xAs_2$ with x = 0, 1, and 2, are grown out of self-flux. For $BaNi_2As_2$, careful single crystal and powder x-ray diffraction results show the low-temperature structure as triclinic. The anisotropic electrical resistivity and magnetic susceptibility results are compared for $BaFe_{2-x}Ni_xAs_2$. The thermodynamic and transport properties of $BaFe_2As_2$ appear to be different from $BaNi_2As_2$. BaFeNiAs$_2$ exhibits properties similar to $BaCo_2As_2$. The experimental details below are followed by a discussion of the room and low temperature crystal structures. The thermodynamic and transport properties will then be presented and discussed for each, sequentially.

## 2. Experimental details

In the preparation of crystals, high purity elements (> 99.9 %) were used; the source of the elements was Alfa Aesar. Large single crystals of $BaFe_2As_2$, $BaNi_2As_2$, and BaFeNiAs$_2$ were grown out of FeAs, NiAs, and FeAs:NiAs binaries, respectively. Self flux is preferred over other metal solvents such as Sn or Pb, as flux impurities can become incorporated in the crystals; this is recognized to be particularly important issue in $BaFe_2As_2$ [26]. FeAs was prepared by placing mixtures of As and Fe pieces in a silica tube. Fe and As pieces were reacted slowly by heating to 700 °C (dwell 6 hrs), then to 1065 °C (dwell 10 hrs). NiAs binary was prepared by placing mixtures of As, and Ni powder in a silica tube. These were reacted slowly by heating to 300 °C (dwell 3 hrs), to 600 °C (dwell 20 hrs), then to 650 °C (dwell 10 hrs). For $BaFe_2As_2$ crystals, a ratio of Ba:FeAs = 1:5 was heated for 8 hours at 1180°C under partial argon atmosphere. The ampoule was cooled at the rate of 4 °C/hour, followed by decanting of FeAs flux at 1090 °C [27]. For the growth of $BaNi_2As_2$ single crystals, a ratio of Ba:NiAs = 1:4 was heated in an alumina crucible for 10 hours at 1180°C under partial atm argon. This reaction was cooled at the rate of 3 °C/hour, followed by decanting of flux at 1025 °C. The BaFeNiAs$_2$ crystals were prepared in a similar way, but using a near equal molar admixture of FeAs and NiAs. The typical crystal sizes from all batches were $\sim 6 \times 5 \times 0.2$ mm$^3$. The crystals were brittle, well-formed plates with the [001] direction perpendicular to the plane of the crystals. The $BaNi_2As_2$ crystals were found to be highly air-sensitive.

Electron probe microanalysis of a cleaved surface of a $BaNi_2As_2$ single crystal was performed on a JEOL JSM-840 scanning electron microscope using an accelerating voltage of 15 kV and a current of 20 nA with an EDAX brand energy-dispersive X-ray spectroscopy (EDS) device. This gave Ba:Ni:As = 1:2:2. EDS analyses on the $BaFeNiAs_2$ crystal indicated Ba:Ni:Fe:As ratio of ~ 1:1:1:2.

Powder x-ray diffraction data at room temperature were collected using PANalytical X'Pert Pro MPD with an X'celerator position sensitive detector (Cu $K_\alpha$ radiation) for $BaFe_2As_2$, $BaNi_2As_2$, and $BaFeNiAs_2$. For $BaNi_2As_2$ the data was also collected at 105 K using an Anton Parr TTK450 low temperature stage. The unit cell parameters were refined from full pattern LaBail fits using the program FullProf [28].

Single crystal x-ray diffraction data on $BaNi_2As_2$ were also collected at 293 K. For room temperature data collection the crystal was mounted inside a 0.3 mm capillary filled with argon. Further details for this procedure can be found in Ref. [29]. For low-temperature data collection, a small single crystal of $BaNi_2As_2$ was mounted on the top of a thin glass capillary using the Fombin Z15 oil in a a stream of nitrogen gas at 173 K. Further details on crystal mounting can be found in Ref. [30]. The single crystal was transferred to a Bruker Apex II diffractometer using cryotongues and placed under the stream of nitrogen gas at 100 K. The single crystal data were collected at 100(1) K, then the temperature of nitrogen gas was raised to 160(1) K and the data were collected again.

DC magnetization was measured as a function of temperature and field using a Quantum Design Magnetic Property Measurement System for $BaFe_2As_2$, $BaNi_2As_2$, and $BaFeNiAs_2$. For a temperature sweep experiment, the sample was cooled to 1.8 K in zero-field (zfc) and data were collected by warming from 1.8 K to 400 K in an applied field of 10 kOe. The magnetic susceptibility results are presented per mole of formula unit ($cm^3$/mol), $\chi$, along *c*- and *ab*-crystallographic directions. The polycrystalline averages of the susceptibilities can be estimated by $\chi = [2\chi_{ab}+\chi_c]/3$.

Temperature dependent electrical resistivity was performed on a Quantum Design Physical Property Measurement System (PPMS). The electrical contacts were placed on samples in standard 4-probe geometry, using Pt wires and silver epoxy (EPO-TEK H20E). The resistivity was measured in the *ab*-plane ($\rho_{ab}$) and *c*-direction ($\rho_c$) for $BaFe_2As_2$, $BaNi_2As_2$, and $BaFeNiAs_2$. Specific heat data, $C_p(T)$, were also obtained using the PPMS via the relaxation method from 200 K to 2 K. For $BaNi_2As_2$ data were collected down to 0.4 K using a $^3He$ insert.

## 3. Results

### 3.1 Crystal Structures

At room temperature, the structures are identified as the tetragonal $ThCr_2Si_2$ (*I*4/*mmm*, Z = 2), based on x-ray powder diffraction data. The lattice constants for $BaFe_2As_2$, $BaFeNiAs_2$, and $BaNi_2As_2$ are shown in Fig. 1, along with data for $BaCo_2As_2$. On moving from Fe to Ni, *a* increases and *c* decreases. This results in an increased distortion of the tetrahedral environment of the transition metal atoms in $BaT_2As_2$. The lattice parameters of $BaFe_2As_2$ are *a* = 3.9635(5) Å and *c* = 13.022 (2) Å, and those of $BaFeNiAs_2$ are $a$ = 4.0002(1) Å and $c$ = 12.6767 (3) Å, similar to those of $BaCo_2As_2$ [31]. The lattice parameters of $BaNi_2As_2$ are $a$ = 4.1441(1) Å and $c$ = 11.6325(3) Å, slightly larger than those recently reported for the Pb-flux grown crystals with $a$ = 4.112(4) Å and $c$ = 11.54(2) Å [18], and in better agreement with the original reports [32].

With cooling, $BaFe_2As_2$ and $BaNi_2As_2$ undergo symmetry-breaking transitions. For $BaFe_2As_2$, a transition below 136 K is associated with a tetragonal to orthorhombic *Fmmm* change in space group. Below the phase transition, the four equal Fe-Fe bonds at 280.2 pm are split into two pairs with 280.8 pm and 287.7 pm lengths [2]. For $BaFeNiAs_2$, the low temperature diffraction was not studied, as no structural transition is expected from the bulk properties (Sec 3.2.3). For $BaNi_2As_2$, a tetragonal to orthorhombic first-order phase transition was suggested below $T_0$ = 130 K only by analogy with $AFe_2As_2$ ($A$ = Ba, Sr, Ca) [22-25] compounds, however, we find that the symmetry of the low temperature structure is triclinic.

Upon cooling below 130 K a structural distortion is observed in $BaNi_2As_2$. Fig. 2a shows two sections of powder x-ray diffraction data collected at 298 K and 105 K, and illustrates the complex diffraction pattern observed at low temperature due to the splitting of the Bragg peaks as the symmetry is reduced. The diffraction pattern at 105 K could only be indexed after removal of all of the symmetry constraints, resulting in a triclinic unit cell. Difficulties related to preferred orientation and the complexity of the low temperature structure of $BaNi_2As_2$ precluded satisfactory refinement of the crystal structure from the powder x-ray diffraction data. In order to identify the low temperature unit cell accurately, single crystal x-ray diffraction was employed.

The single crystal x-ray diffraction data for $BaNi_2As_2$ were collected at 100 K. Due to the symmetry breaking, the low temperature crystal was twinned, requiring a careful symmetry analysis before any structural solution and refinement was possible. Because the unit cell distortion is small and the resolution of the CCD detector is relatively poor, the spots resulting from different twin components could not be spatially resolved for unit cell determination or intensity integration. Still, indexing of the Bragg reflections suggested a possible monoclinic ($\alpha$ = ~89°) or even triclinic symmetry. Because the x-ray powder diffraction data is immune to twinning problems and provides a larger angular resolution, it was employed to differentiate between the two cells. As described above, the multiple peaks splitting of the powder pattern at 105 K could be indexed only within the triclinic cell.

The structural solution and refinement of the single crystal x-ray data were undertaken both in the $P1$ and $P\bar{1}$ space groups. While the $P1$ space group provided a reasonable solution, strong correlations between different sites containing the same atom types pointed to a higher symmetry. Thus, the structural solution and refinement proceeded in the $P\bar{1}$ space group. A numerical absorption correction was based on the crystal shape derived from the optical face indexing. The twinning law accounting for the tetragonal-type twinning (rotation by 90° around the original tetragonal axis) was employed during the final stages of refinement. The refined fraction of the major twin component was 87%, while the scattering contributions of the other three components were 2 and 3 and 8%. The structural parameters at 293 K and 100 K, from single crystal x-ray diffraction, are listed in Table 1. The resulting triclinic unit cell is outlined in Fig. 2b, and can be visualized as a distortion of the primitive unit cell of the body centered $ThCr_2Si_2$ structure. The powder diffraction pattern at 105 K was also indexed to a similar triclinic cell (Fig. 2a), with *a* = 6.4979(9) Å, *b* = 6.4911(9) Å, *c* = 6.4444(9) Å, α =37.30(1)°, β = 54.36(1)° , γ = 37.39(1)°. After heating to 160 K, the single crystal x-ray data could be indexed in the tetragonal unit cell. The peaks in the 298 K powder diffraction data were also indexed to the tetragonal cell (Fig. 2a).

At room temperature the transition metal atoms lie on a perfect square net in the *ab*-plane of the tetragonal structure, and for $BaNi_2As_2$, the Ni-Ni distance is simply $a/2^{1/2}$ = 2.9327(6) Å. In $BaNi_2As_2$, the further reduction in symmetry below $T_0$ results in a distorted Ni network as shown in Fig. 2c. The Ni atoms still lie nearly in a plane, but within the plane form zigzag chains with short Ni-Ni contacts (2.8 Å). The chains are separated by significantly longer Ni-Ni distances (3.1 Å).

Finally, we note that reciprocal space analysis of the 100 K single crystal x-ray diffraction indicated tripling of one of the directions (the *b* direction within the tetragonal setting), shown in Fig. 2d. The superstructure reflections were very weak, with their intensities 2 to 3 times the background level. Because of the low intensities and limited number of the superstructure reflections, no reliable structural solution could be obtained. While the tetragonal-type twinning would require the presence of the superstructure reflections along the other tetragonal direction, the weak nature of the superstructure reflections and dominance of one twin component are the most likely reasons that no superstructure reflections were observed along another direction.

## 3.2 Physical properties

### 3.2.1 $BaFe_2As_2$

For $BaFe_2As_2$, the magnetic susceptibility at room temperature gives $\chi_c \approx \chi_{ab} \approx 6.7 \times 10^{-4}$ cm$^3$ mol$^{-1}$ (Fig. 3a). The susceptibility decreases linearly with decreasing temperature, and drops abruptly below ~ 135 K with $\chi_c > \chi_{ab}$ below. As seen in the bottom inset of Fig. 3a, the field dependent magnetization is linear at 1.8 K. For this compound, the polycrystalline average of the susceptibility data are presented as the Fisher's $d(\chi T)/dT$ [33] versus temperature, to infer an antiferromagnetic transition temperature, at $T_N$ = 132 K (the top inset of Fig. 3a).

Temperature dependent electrical resistivity for $BaFe_2As_2$ decreases with decreasing temperature, supporting metallic behavior (Fig. 3b). The resistivity behavior is highly anisotropic with $\rho_c > \rho_{ab}$. At room temperature, $\rho_{ab}$ = 0.50 mΩ cm and $\rho_c$ = 16.44 mΩ cm for $BaFe_2As_2$. The residual resistivity ratio RRR (= $\rho_{300K}/\rho_{2K}$) is 4 along both directions, indicative of decent crystal quality. Below ~135 K and for $\rho_c$ and $\rho_{ab}$ there is a sharp step-like drop. This feature is best manifested in the derivative of resistivity, $d\rho/dT$, giving a peak at 132(1) K (bottom inset of Fig. 3b). The magnitude of the resistivity at 2 K and 8 Tesla is higher than the zero-field value for $BaFe_2As_2$ (top inset of Fig. 3b). This corresponds to a positive magnetoresistance, $(\rho_{8T}-\rho_0)/\rho_0$, of 27.7 % at 2 K. The magnetoresistance for $\rho_c$ is slightly negative and -4.6%, at 2 K.

Fig. 3c gives the temperature dependence of specific heat. For $BaFe_2As_2$, specific heat gives a pronounced lambda anomaly, peaking at 132(1) K. Below ~ 6 K, C/T has a linear $T^2$ dependence (bottom inset of Fig. 3c). The fitted Sommerfeld coefficient, $\gamma$, for $BaFe_2As_2$ is 6.1 mJ/($K^2$mol) [or 3.0 mJ/($K^2$mol Fe)]. The Debye temperature is $\theta_D \approx 260$ K estimated above 150 K, using the calculated values of the $T/\theta_D$ dependence of the Debye specific heat model. Previously, values of $\gamma = 16(2)$ mJ/($K^2$mol) and $\theta_D = 134(1)$ K were reported from fits between 3.1 and 14 K for $BaFe_2As_2$ [2].

$BaFe_2As_2$ properties (Fig. 3) demonstrate the consistency among of C(T), $d(\chi T)/dT$ and $d\rho/dT$ in determining $T_N = 132(1)$ K. This transition temperature is lower than that previously reported at 140 K [2].

### 3.2.2 $BaNi_2As_2$

For $BaNi_2As_2$, the magnetic susceptibility is anisotropic with $\chi_{ab}/\chi_c \approx 1.6$ at 10 kOe (Fig. 4a). $\chi$ is roughly temperature independent and presents only a small drop at $T_0 \approx 132$ K and a rise below ~ 12 K. The field dependent magnetization is linear at 1.8 K along *c*- and *ab*- crystallographic directions (Fig. 4a, inset).

$BaNi_2As_2$ shows metallic behavior (Fig. 4b) with $\rho_c > \rho_{ab}$. At room temperature, $\rho_{ab} = 0.07$ mΩ cm and $\rho_c = 4.69$ mΩ cm. Along *c* and below $T_0$, resistivity abruptly drops two orders of magnitude. Along *ab* there is a sharp increase at $T_0$ in $\rho$ followed by a continuous decrease with decreasing temperature. The residual-resistivity ratio RRR (= $\rho_{300K}/\rho_{2K}$) is 8 along *ab* illustrating a better quality crystal compared to those grown out of Pb [19]. The superconductivity downturn in resistivity starts around 1.5 K in our pure crystal (inset of Fig. 4b), similar to the recently reported [19]. $T_c$ as determined by the zero resistance state is 0.69 K, comparable to the crystals grown in Pb-flux [19]. There is no noticeable magnetoresistance at 8 Tesla for $BaNi_2As_2$ at 1.8 K (data not shown).

Fig. 4c gives the temperature dependence of specific heat for $BaNi_2As_2$. There is a sharp, pronounced transition at $T_0 = 132$ K with a width of less than 0.2 K, in contrast to the broader, second-order-like transition observed in $BaFe_2As_2$. Between 1.8 K and ~ 6 K, the normal state electronic specific heat coefficient, $\gamma_n$, is 13.2 mJ/($K^2$mol) [or 6.6 mJ/($K^2$mol Ni)]. The Debye temperature is estimated as $\theta_D \approx 250$ K. The value of $\gamma_n$ is larger than the corresponding value for $LaNiAsO_{0.9}F_{0.1}$, $\gamma_n = 4.75$ mJ/($K^2$mol) [17], and comparable to other layered superconductors $Li_xNbS_2$ with $\gamma_n = 10$ mJ/($K^2$mol) [34] and $NaCoO_2$ $\gamma_n = 24$ mJ/($K^2$mol) [35, 36]. Previous measurements of Pb-flux grown crystal gave $\gamma_n = 10.8(1)$ mJ/($K^2$mol) and $\theta_D = 206$ K [19].

For $BaNi_2As_2$, the temperature dependence of C/T in several fields for H//*ab* is shown in Fig. 5. In zero-field, there is a sharp anomaly at $T_c = 0.69$ K with a jump $\Delta C = 12.6$ mJ/(K mol). Taking the value of the Sommerfeld coefficient at $T_c$, this gives a ratio of $\Delta C/\gamma_n T_c = 1.38$, roughly comparable to a value of 1.43 predicted by weak-coupling BCS theory and confirms bulk superconductivity. As the magnetic field increases, the transition becomes broader and shifts to lower temperatures. The upper critical field $H_{c2}$ is determined by onset criterion, demonstrated by the dashed line, and displayed in the inset of Fig. 5. The solid line is a linear fit to the data, yielding a slope of $-dH_{c2}/dT_c$ = - 0.142 T/K. Using the Werthamer-Helfand-Hohenberg equation $H_{c2}(0) = -0.693 T_c(dH_{c2}/dT_c)$ [37], this gives $H_{c2}(0) = 0.069$ Tesla. A zero temperature, coherence length of $\xi_{GL}(0) \approx 691$ Å is obtained using Ginzburg-Landau coherence length formula $\xi_{GL}= (\Phi_0/2\pi H_{c2}(0))^{½}$, where $\Phi_0 = 2.07 \times 10^{-7}$ Oe cm$^2$.

### 3.2.3 $BaFeNiAs_2$

The temperature dependent magnetic susceptibility at 10 kOe for $BaFeNiAs_2$ is found to be essentially temperature independent down to 25 K, rapidly increasing with further decreasing of temperature (Fig. 6a). The averaged polycrystalline magnetic susceptibility at 1.8 K is $\chi = 2.3 \times 10^{-3}$ cm$^3$ mol$^{-1}$ and at room temperature is $9.3 \times 10^{-4}$ cm$^3$ mol$^{-1}$. These values are comparable to that for $BaCo_2As_2$ [31].

Temperature dependent electrical resistivity gives metallic behavior for $BaFeNiAs_2$ (Fig. 6b). At room temperature, $\rho_{ab} = 0.08$ mΩ cm and $\rho_c = 2.25$ mΩ cm. The residual resistivity ratio RRR is 12 along *ab*, but only ~ 2 along *c*. There is no noticeable magnetoresistance in 8 Tesla (data not shown). For $BaFeNiAs_2$ the resistivity data exhibit a quadratic temperature dependence, $\rho = \rho_o + AT^2$, over a wide temperature range of 1.8 K to ~ 68 K. The *A* value is a measure of the extent of electron correlations. Plotted in the inset of Fig. 6b is $\rho(T^2)$, and linear fits in this region. For $\rho_{ab}$, $A = 3.1 \times 10^{-6}$ mΩ cm K$^{-2}$, and it is slightly larger than for $BaCo_2As_2$ at $2.2 \times 10^{-6}$ mΩ cm K$^{-2}$ [31], consistent with the formation of a Fermi-liquid state.

Fig. 6c gives the temperature dependence of specific heat results. For $BaFeNiAs_2$, there are no features in specific heat up to 200 K, similar to that for $BaCo_2As_2$. The inset of Fig. 5b shows C/T versus $T^2$ dependence. $BaFeNiAs_2$ gives a slight upturn below ~ 4 K; the $\gamma$ is remarkably similar to those for $BaCo_2As_2$. For $BaCo_2As_2$, C/T vs. $T^2$ is linear below ~ 8 K consistent with a Fermi liquid plus phonon contribution. $BaCo_2As_2$ gives $\gamma$ = 41.6 mJ/(K$^2$mol) [or 20.8 mJ/(K$^2$mol Co)]. The Wilson ratio, $R_w = \pi^2 k_B^2\chi/(3\mu_B^2\gamma)$, for $BaCo_2As_2$ is ~ 7 from $\chi_{ab}$ and ~ 10 from $\chi_c$ [31]. These values well exceed unity for a free electron system and indicate that the system is close to ferromagnetism.

## 4. Final Discussions

The shrinking of the crystal lattice is observed from $BaFe_2As_2$, to $BaFeNiAs_2$, and finally to $BaNi_2As_2$. The cell volumes at room temperature are 204.567(2) Å$^3$, 202.8517(3) Å$^3$, and 199.7686(2) Å$^3$, respectively. With cooling, $BaFe_2As_2$ and $BaNi_2As_2$ undergo structural transitions. For $BaFe_2As_2$, a second-order-like transition is observed at 132(1) K, associated with a tetragonal to orthorhombic *Fmmm* distortion, and a SDW magnetic transition. In contrast to this broader transition, for $BaNi_2As_2$ there is sharp, pronounced feature at $T_0$ = 131 K, seen in $\rho(T)$ and $C_p$ results. This first order transition was associated with a reduction of the lattice symmetry from tetragonal to triclinic $P\bar{1}$. The most interesting effect of the structural distortions in these two compounds is the resulting change in the Fe or Ni network. At room temperature the transition metal atoms lie on a perfect square net in the *ab*-plane of the tetragonal structure. In $BaFe_2As_2$, the low temperature phase is orthorhombic and the Fe network is distorted from square to rectangular, commensurate with magnetic ordering. In $BaNi_2As_2$, the further reduction in symmetry below $T_0$ results in a more distorted Ni network, and zigzag chains of Ni atoms. $BaFeNiAs_2$ shows no evidence of structural or magnetic transitions.

The basic thermodynamic and transport properties of pure $BaFe_2As_2$, $BaNi_2As_2$, and $BaNiFeAs_2$ crystals show differences between them. Unlike $BaFe_2As_2$, the magnetic susceptibility for $BaNi_2As_2$ is roughly temperature independent and presents only a small anomaly at $T_0$. At room temperature, $\chi_{avg}$ ($\times 10^{-4}$ cm$^3$ mol$^{-1}$) is smallest for $BaNi_2As_2$ at 3, and slightly larger for $BaFe_2As_2$ at 7 and largest for $BaFeNiAs_2$ at 9. $BaNi_2As_2$ is more conductive at room temperature, compared to $BaFe_2As_2$. The room temperature resistivity for $BaFeNiAs_2$ and $BaNi_2As_2$ are at ~ 0.08 mΩ cm, compared to $\rho_{ab}$ = 0.50 mΩ cm for $BaFe_2As_2$. At room temperature, $\rho_c$ = 16.44 mΩ cm for $BaFe_2As_2$, approximately one order of magnitude larger than for $BaFeNiAs_2$ and $BaNi_2As_2$.

Superconductivity in $BaNi_2As_2$ is connected to pure sample and those lightly doped with Fe. The $BaFe_{2-x}Ni_xAs_2$ series of $0 \leq x \leq 0.2$ gives the highest $T_c^{onset}$ = 21 K for x = 0.1 [38]. From temperature dependent resistivity measurements, we find no sign of superconductivity down to 0.4 K for x = 0.4, 0.6, 0.8 and 1. The results of $BaFeNiAs_2$ was given here, as a half point between $BaNi_2As_2$ ($d^8$) and $BaFe_2As_2$ ($d^6$). $BaFeNiAs_2$ is found to show a similar magnitude of magnetic susceptibility as well as specific heat $\gamma$ to $BaCo_2As_2$, which was suggested to be in a close proximity to a quantum critical point and of strong fluctuations [31]. The calculated local density approximations show hybridization between As *p* states and Fe *d* states in $BaFe_2As_2$ [39] similar to $BaNi_2As_2$ [40] and $BaCo_2As_2$ [31]. However the Fermi level for $BaNi_2As_2$ is shifted up because of the higher valence electron count in $Ni^{2+}$ ($3d^8$) compared to $Fe^{2+}$ ($3d^6$). With a larger Fermi surface, $BaNi_2As_2$ shows large Fermi surfaces and lower density of states [13, 40]. Based on band structure calculations, the superconductivity of $BaNi_2As_2$ appears to be rather different from $BaFe_2As_2$. The Ni-phase can be understood within the context of conventional electron-phonon theory, while the high $T_c$ Fe superconductors can not be described in this way.

## Acknowledgement

Research sponsored by the Division of Materials Science and Engineering, Office of Basic Energy Sciences. Part of this research was performed by Eugene P. Wigner Fellows at ORNL. Work at Los Alamos was performed under the auspices of the U. S. Department of Energy.

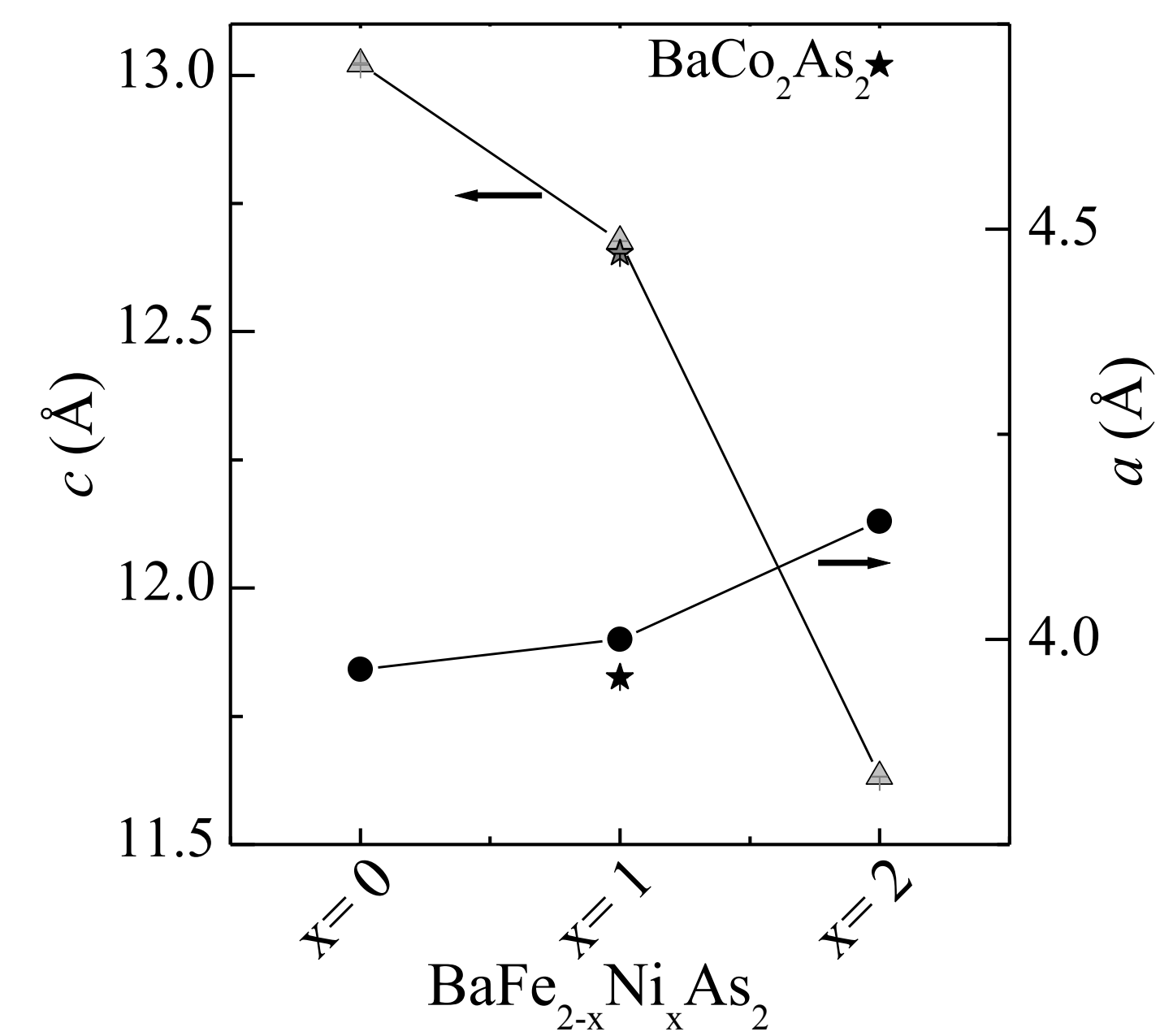

Fig. 1: Room temperature lattice constants for $BaFe_2As_2$, $BaFeNiAs_2$, and $BaNi_2As_2$, refined from x-ray powder diffraction data; those for $BaCo_2As_2$ are shown as stars.

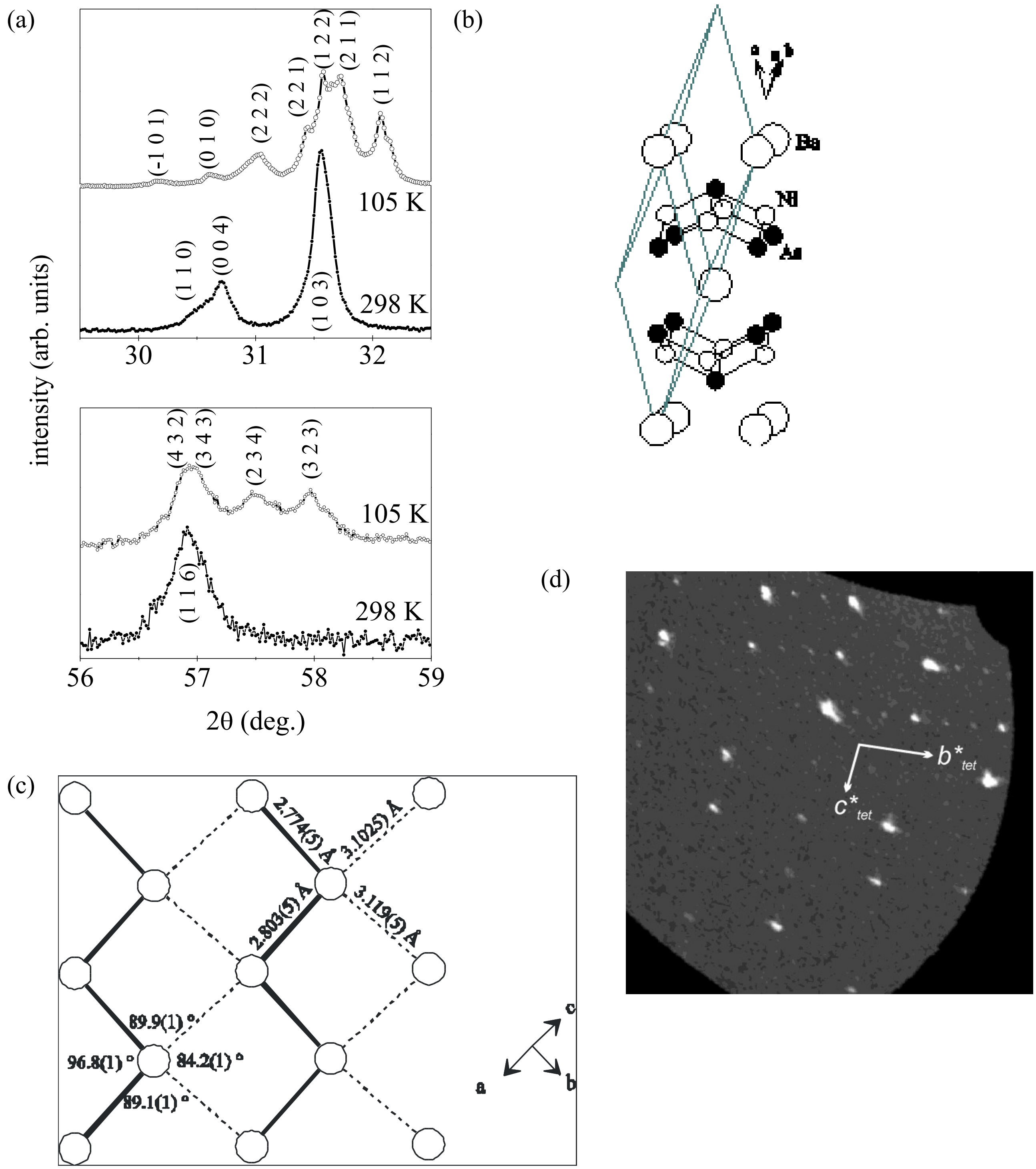

Fig. 2: (a) For $BaNi_2As_2$, two sets of Bragg peaks from powder x-ray diffraction data at 298 K and 105 K. (b) The low-temperature triclinic symmetry for $BaNi_2As_2$; the unit cell is represented with lines. (c) The distorted Ni-network for low-temperature $BaNi_2As_2$ structure. The Ni atoms form zigzag chains with short and longer Ni-Ni bonds within the plane. (d) The image of the reciprocal space for $BaNi_2As_2$ at 100 K indicates weak superstructure reflections, tripling along the *b* direction.

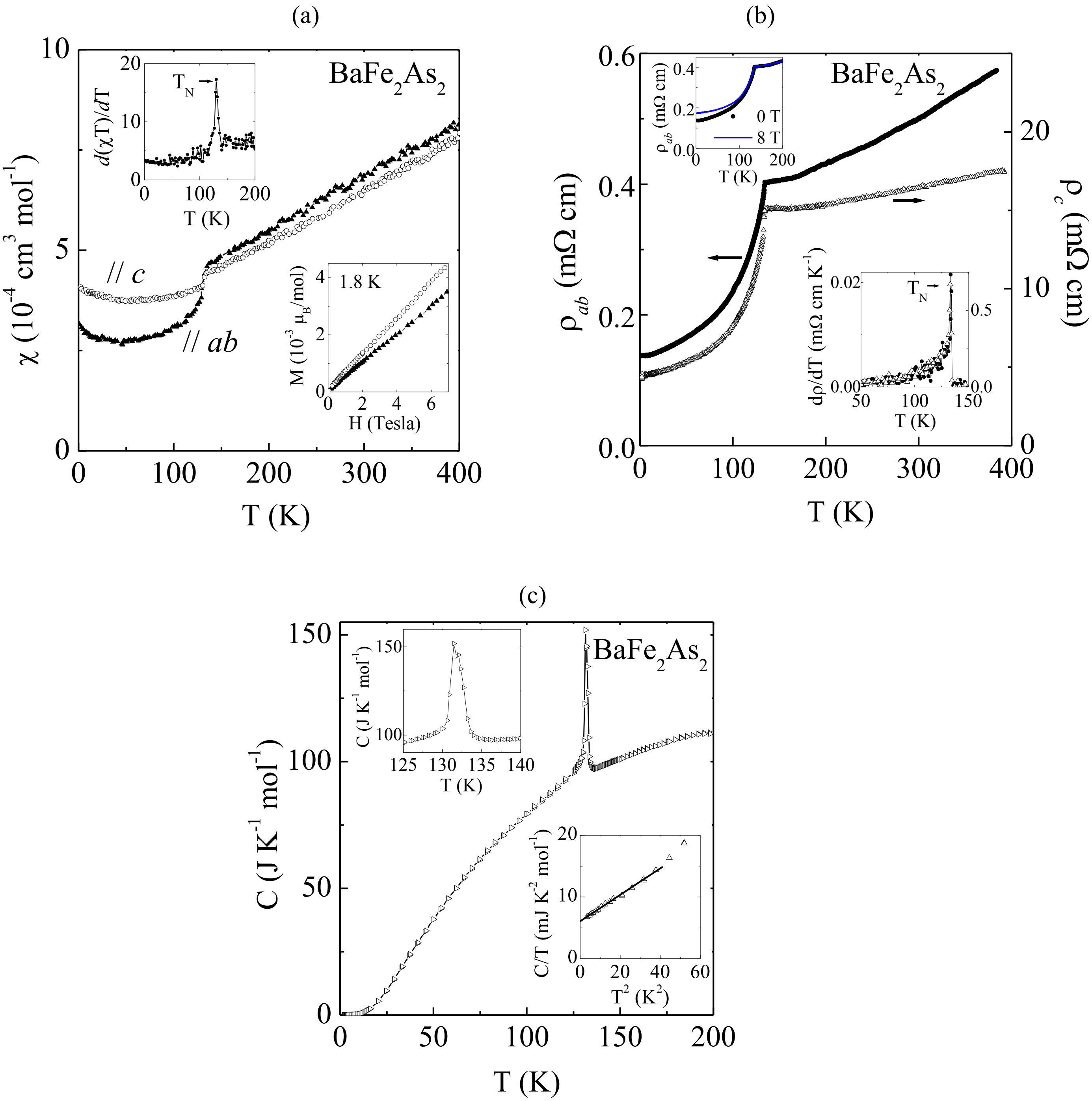

Fig. 3: For $BaFe_2As_2$ and along the two crystallographic directions, the temperature dependence of (a) molar susceptibility in zero-field cooled forms measured at 10 kOe, and (b) resistivity at zero field. (c) For $BaFe_2As_2$, the temperature dependence of the specific heat between 1.8 K and 200 K. The inset of (a) shows the field dependence of magnetization and the average susceptibility in the form of $d(\chi T)/dT$ peaking at $T_N$. Inset of (b) is the derivative in resistivity at zero field, $d\rho/dT$, peaking at $T_N$; also $\rho_{ab}(T)$ at 0 and 8 Tesla. Inset of (c) is the C/T vs. $T^2$ and a linear fit between 1.8 K and ~ 7 K.

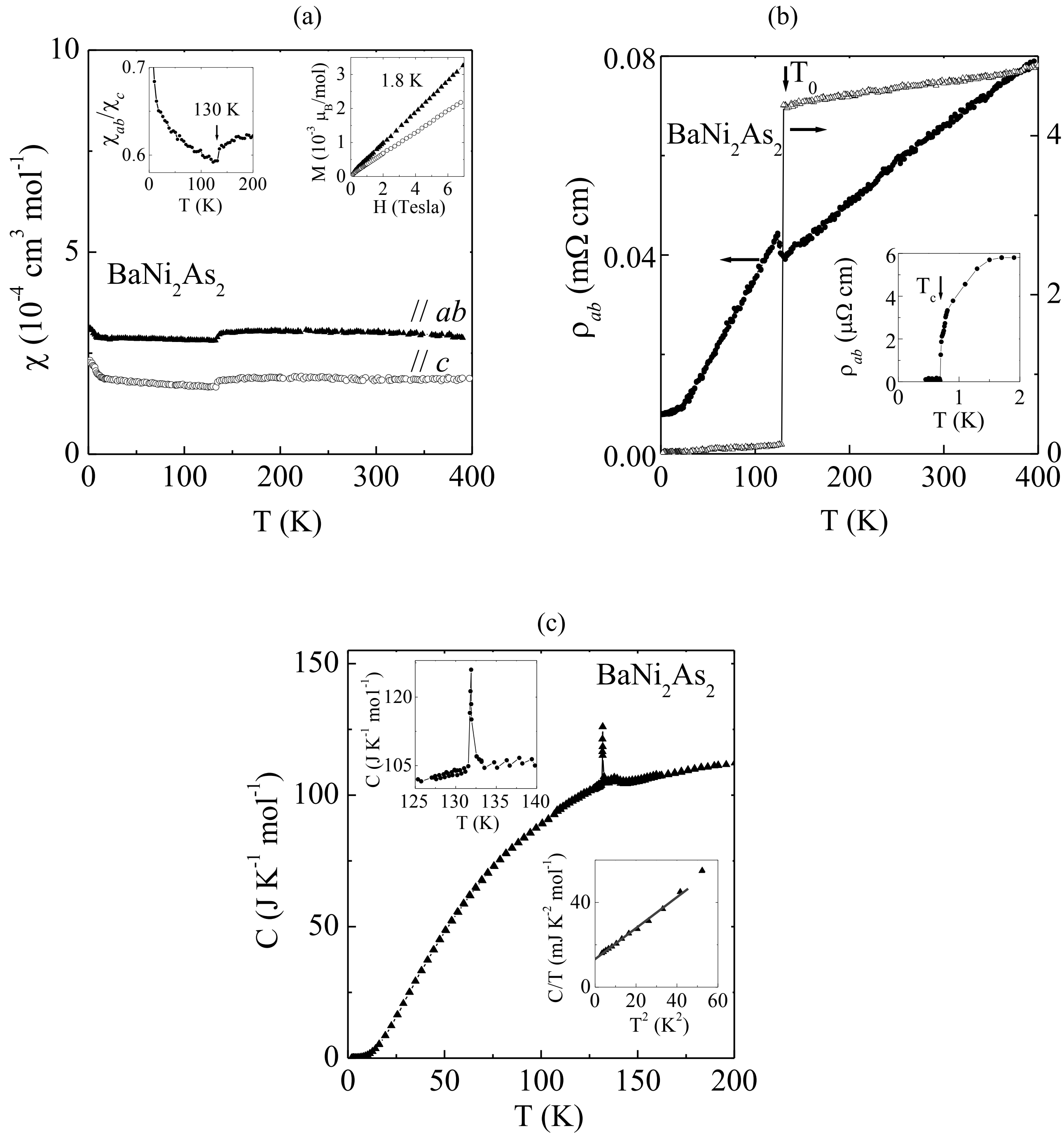

Fig. 4: For $BaNi_2As_2$ and along the two crystallographic directions, the temperature dependence of (a) molar susceptibility in zero-field cooled forms measured at 10 kOe, and (b) resistivity at zero field. (c) For $BaNi_2As_2$, the temperature dependence of specific heat between 1.8 K and 200 K. The inset of (a) shows the field dependence of magnetization and $\chi_{ab}/\chi_c$. Inset of (b) is the plot of $\rho_{ab}(T)$ below 2 K, marking the critical temperature. Inset of (c) is the C/T vs. $T^2$ and a linear fit between 1.8 K and ~ 7 K.

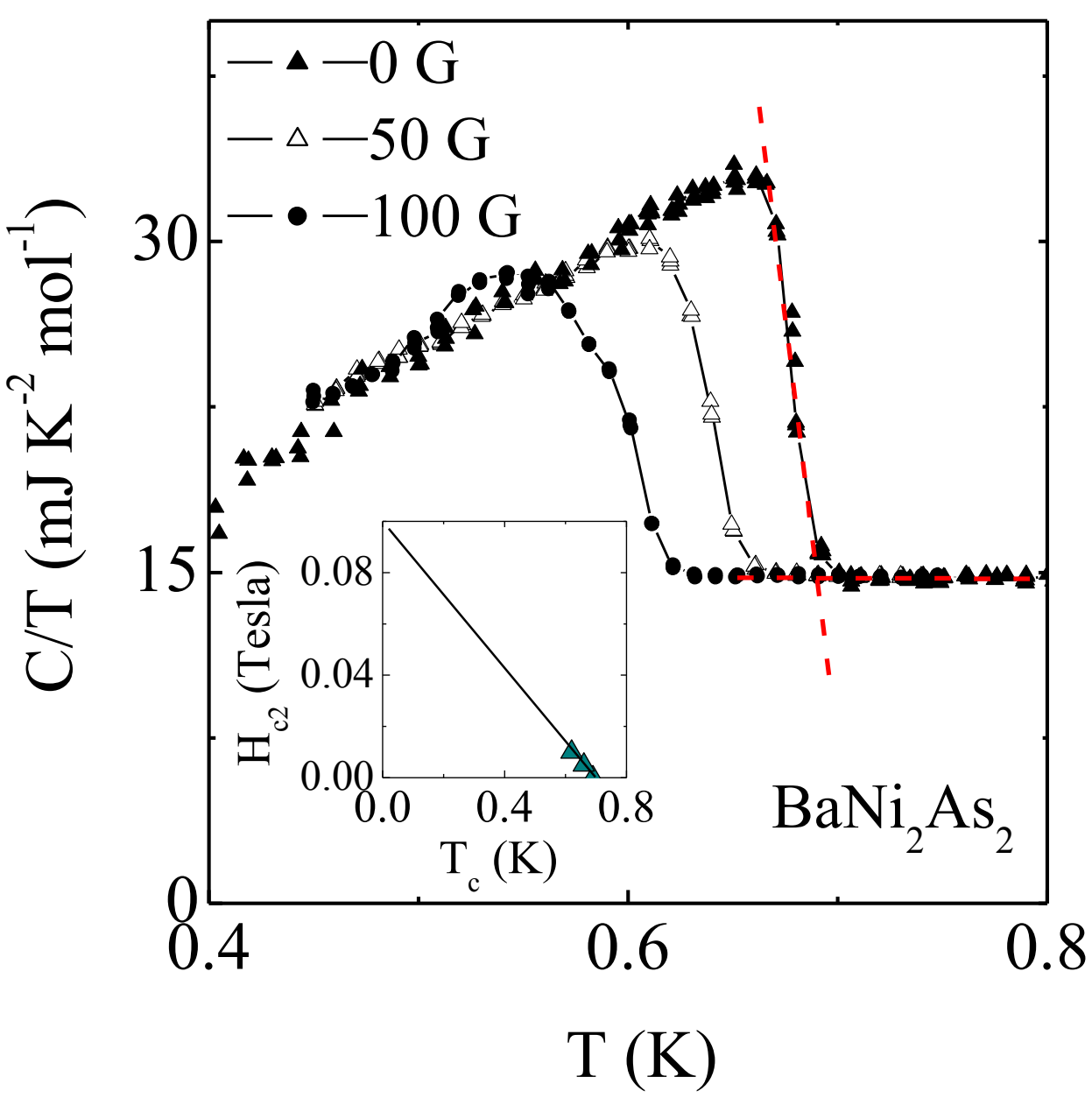

Fig. 5: For $BaNi_2As_2$, the field dependence of specific heat below 0.8 K. The $T_c$ was found using the onset criterion demonstrated by the dashed line for the zero-field data. The inset shows the upper critical field $H_{c2}$ for H//*ab* assigned using the onset criterion, with the solid line representing a least squares fit to the data.

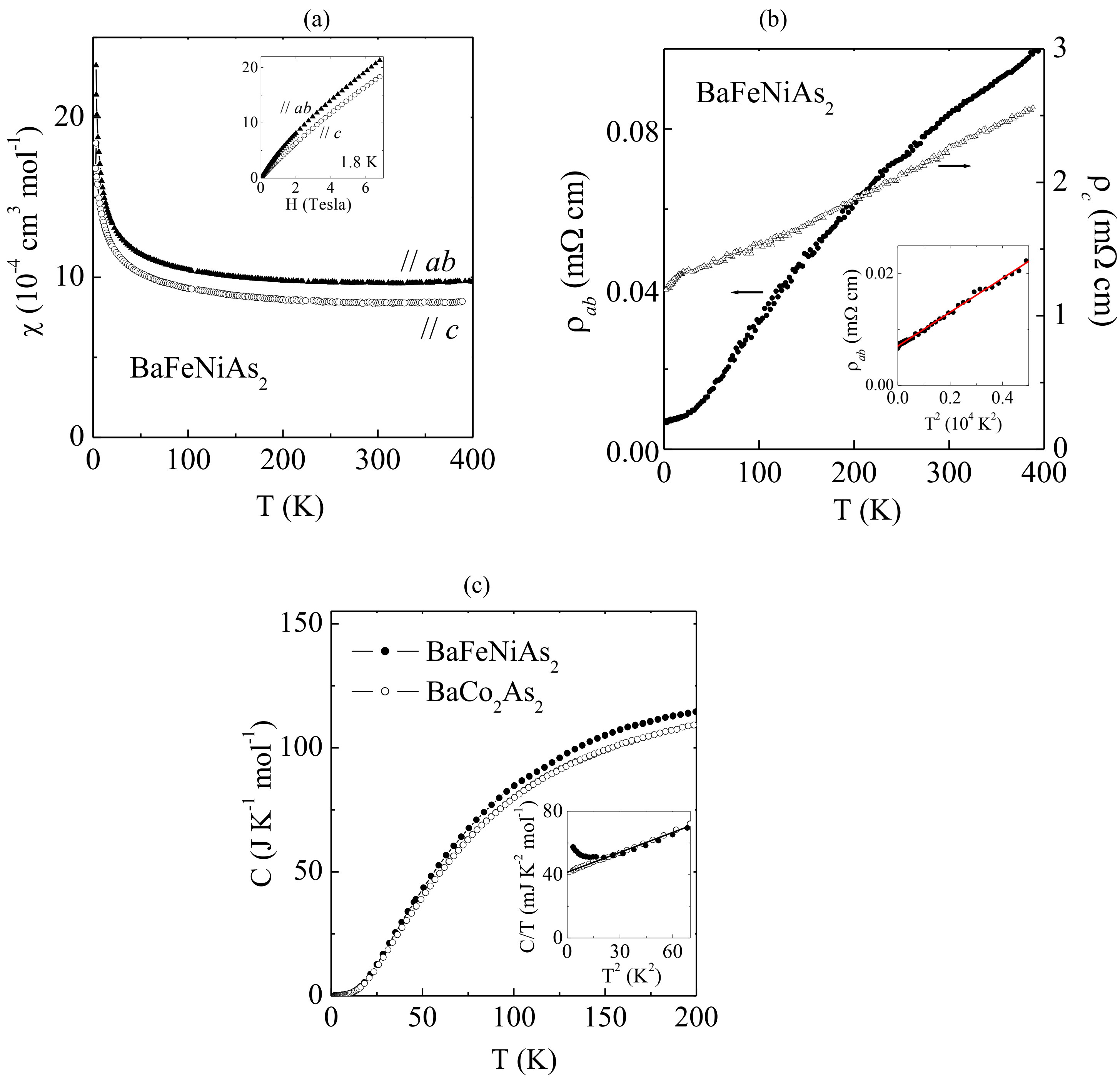

Fig. 6: For $BaFeNiAs_2$ and along the two crystallographic directions, the temperature dependence of (a) molar susceptibility in zero-field cooled forms measured at 10 kOe, and (b) resistivity at zero field. (c) For $BaFeNiAs_2$ and $BaCo_2As_2$, the temperature dependence of the specific heat between 1.8 K and 200 K. The inset of (a) shows the field dependence of magnetization at 1.8 K. Inset of (b) is the plot of $\rho_{ab}$ vs. $T^2$ and linear fit between 2 and 68 K. The inset of (c) is the C/T vs. $T^2$ and a linear fit between 1.8 and 7 K for $BaCo_2As_2$.

Table 1: Single crystal data and structure parameters for $BaNi_2As_2$, at 100 K and 293 K. The data was collected at λ = 0.71073 Å.

| | | |
|---|---|---|
| Temperature (K) | 100 | ***293*** |
| Crystal system | Triclinic | Tetragonal |
| Space group | $P\bar{1}$ | $I4/mmm$ |
| $Z$ | 1 | 2 |
| Unit cell dimensions (Å) | $a$ = 6.5170(14), α =37.719(11)°<br>$b$ = 6.4587(12), β = 54.009(14)°<br>$c$ = 6.4440(12), γ = 37.382(9)° | $a$ = $b$ = 4.1474(9)<br>$c$ = 11.619(2) |
| Volume (Å$^3$) | 100.13(3) | 199.85(7) |
| 2θ range (°) | 3.53 to 32.97° | 3.51 to 28.62° |
| Reflections collected | 1567 | 802 |
| Conventional residual, $R_1$,<br>Weighted residual, $wR_2$, [I >2σ(I)]. | 0.0682<br>0.1836 | 0.0230<br>0.0430 |
| Largest diff. peak, hole (e Å$^{-3}$) | 3.550, -4.076 | 0.985, -1.365 |
| Atomic coordinates (*Wyckoff*): | Ba (1*a*): 0,0,0<br>As (2*i*):0.3669(3), 0.9785(5), 0.3444(5)<br>Ni (2*i*): 0.2219(4), 0.5255(7), 0.7513(7) | Ba (2*a*): 0,0,0<br>As (4*e*): 0, 0, 0.3471(1)<br>Ni(4*d*): 0, 0.5, 0.25 |
| Isotropic displacement (Å$^2$) Ba | 0.0131(2) | 0.0155(3) |
| As | 0.0277(3) | 0.0197(4) |
| Ni | 0.0366(4) | 0.0221(4) |